\documentclass[sigconf]{acmart}

\usepackage{lipsum}
\usepackage{packages}
\usepackage{float}
\usepackage{rotating}
\usepackage{tabularx}
\usepackage{array}
\usepackage{longtable}
\usepackage{booktabs}   
\usepackage{colortbl}   

\usepackage{float}

\newcolumntype{Y}{>{\centering\arraybackslash}X}

\AtBeginDocument{%
  }

\setcopyright{acmlicensed}
\copyrightyear{2018}
\acmYear{2018}
\acmDOI{XXXXXXX.XXXXXXX}

\acmConference[MSR'26]{IEEE/ACM International Conference on Mining Software Repositories}{April 13-14, 2026}{Rio de Janeiro, Brazil}
\acmISBN{978-1-4503-XXXX-X/2018/06}

\usepackage{xcolor}
\usepackage{comment}
\usepackage{soul}
\usepackage{ulem}
\newboolean{showcomments}
\setboolean{showcomments}{true} 

\ifthenelse{\boolean{showcomments}}
{

}

\newcommand{\Foutse}[1]{{\color{WildStrawberry}{[Foutse: #1]}}}

\begin{document}

\title{Tracing Stereotypes in Pre-trained Transformers:\\From Biased Neurons to Fairer Models}

\author{Gianmario Voria}
\email{gvoria@unisa.it}
\orcid{0009-0002-5394-8148}
\affiliation{%
  \institution{University of Salerno}
  \city{Fisciano}
  \country{Italy}
}

\author{Moses Openja}
\email{openja.moses@polymtl.ca}
\orcid{---}
\affiliation{%
  \institution{Polytechnique Montréal}
  \city{Montréal}
  \country{Canada}
}

\author{Foutse Khomh}
\email{foutse.khomh@polymtl.ca}
\orcid{---}
\affiliation{%
  \institution{Polytechnique Montréal}
  \city{Montréal}
  \country{Canada}
}

\author{Gemma Catolino}
\email{gcatolino@unisa.it}
\orcid{0000-0002-4689-3401}
\affiliation{%
  \institution{University of Salerno}
  \city{Fisciano}
  \country{Italy}
}

\author{Fabio Palomba}
\email{fpalomba@unisa.it}
\orcid{0000-0001-9337-5116}
\affiliation{%
  \institution{University of Salerno}
  \city{Fisciano}
  \country{Italy}
}

\begin{abstract}
The advent of transformer-based language models has reshaped how AI systems process and generate text. In software engineering (SE), these models now support diverse activities, accelerating automation and decision-making.
Yet, evidence shows that these models can reproduce or amplify social biases, raising fairness concerns. Recent work on neuron editing has shown that internal activations in pre-trained transformers can be traced and modified to alter model behavior. Building on the concept of \textit{knowledge neurons}---neurons that encode factual information---we hypothesize the existence of \textit{biased neurons} that capture stereotypical associations within pre-trained transformers.  

To test this hypothesis, we build a dataset of \textit{biased relations}, i.e., triplets encoding stereotypes across nine bias types, and adapt neuron attribution strategies to trace and suppress biased neurons in \textit{BERT} models. We then assess the impact of suppression on SE tasks. Our findings show that biased knowledge is localized within small neuron subsets, and suppressing them substantially reduces bias with minimal performance loss. This demonstrates that bias in transformers can be traced and mitigated at the neuron level, offering an interpretable approach to fairness in SE.

\end{abstract}





\maketitle

\section{Introduction}

The widespread success of deep learning, and particularly the emergence of transformer architectures~\cite{vaswani2017attention}, has enabled \textit{language models (LMs)}, such as BERT~\cite{devlin2019bert} and GPT~\cite{openai2024gpt4ocard}, to become core components of modern AI-enabled systems. These models capture complex linguistic patterns through large-scale pretraining and now underpin a wide range of applications—from healthcare and finance to education and software engineering (SE)~\cite{pena2025evaluating}.  
In SE, transformers have been integrated into tasks such as requirements classification~\cite{voria2025recover}, sentiment and issue analysis~\cite{zhang2025revisiting}, code review~\cite{sun2025bitsai}, and documentation generation~\cite{fukuda2025development}, substantially improving productivity and automation capabilities.

However, the growing use of LMs in socially situated contexts has raised serious concerns about \textit{fairness}---a non-functional requirement increasingly recognized as essential in AI-enabled systems. Pre-trained transformers often reproduce or amplify stereotypes related to gender, race, age, or disability~\cite{mamta2024biaswipe,xu2025biasedit}, which can result in unfair or exclusionary outcomes in socio-technical environments such as developer hiring~\cite{nakano2024nigerian} or team composition~\cite{treude2023she}.  

To address such issues, researchers have proposed a wide range of bias \textit{measurement}, \textit{evaluation}, and \textit{mitigation} techniques~\cite{gallegos2024bias}. Among these, an emerging line of work explores the \textit{structural manipulation of neural networks}, or \textit{neuron editing}, as a means to identify, isolate, and remove biased internal representations~\cite{openja2025fairflrep}.  
These approaches, often referred to as \textit{model surgery}, seek to localize the internal sources of bias, enabling fine-grained interventions that go beyond data or output corrections.  
Yet, prior work has examined bias only at a superficial level, limited both in \textit{depth}, by focusing on broad layer- or pattern-level analyses rather than specific neuron activations, and in \textit{breadth}, by addressing only a few categories (e.g., gender or toxicity) and neglecting the impact of interventions on other model behaviors~\cite{yu2025understanding,xu2025biasedit}.
Consequently, it remains unclear whether bias in transformers is encoded in specific neurons, whether suppressing them reduces stereotypes, and whether suppression degrades performance in SE tasks.

Drawing inspiration from research on \textit{knowledge neurons} \cite{dai2022knowledge}, i.e., neurons shown to encode specific factual associations within transformer feed-forward layers, we hypothesize that biased knowledge is similarly stored in pre-trained transformers, encoded in small, specialized subsets of neurons responsible for activating particular relational associations.  
If such \textbf{biased neurons} exist, identifying and suppressing them could provide a principled and interpretable way to mitigate bias without compromising task performance.  

\steObjectiveBox{\faBullseye\ \textbf{Main Objective.}}{In light of the previous considerations, the objective of this study is to understand the extent to which biased neurons can be traced and suppressed in pre-trained transformers, assessing whether suppression negatively affects downstream performance in software engineering tasks.}

To this aim, we designed an empirical study comprising three main steps.  
First, we construct a dataset of \textit{biased relations}, i.e., triplets encoding stereotypical associations across nine social dimensions, and transform them into bias-activating prompts. This dataset parallels the concept of factual knowledge and aligns with the methodology used for knowledge neuron identification~\cite{dai2022knowledge}.  
Second, we apply neuron attribution and refining strategies~\cite{dai2022knowledge} to pre-trained \textit{BERT}-based models---on which the original knowledge neuron framework was designed, relying on BERT’s bidirectional encoding and cloze-style evaluation (further info in \autoref{sec:models})---to trace biased neurons, and perform targeted \textit{neuron suppression} to measure its effects on bias expression and model perplexity.
Lastly, we evaluate the impact of suppression on fairness-sensitive, non-code software engineering tasks (i.e., incivility, tone bearing, sentiment, and requirements classification), assessing the trade-off between bias mitigation and task performance.

Our results show that biased knowledge in transformers is not diffusely distributed but localized within small subsets of neurons whose suppression markedly reduces stereotypical associations.  
Neuron suppression has only a mild and often negligible impact on model utility across SE tasks, indicating that fairness improvements can be achieved without compromising effectiveness.  
These findings highlight neuron-level tracing as a viable and interpretable approach to understanding and mitigating bias. While our method was evaluated on encoder-based models such as \textit{BERT}, its principles can extend to other transformer architectures, offering a foundation for neuron-level fairness control and transparent model debugging.

\textbf{Our Contribution.} This paper makes three main contributions.  
We first introduce the concept of \textbf{biased relations}, a structured representation of stereotypes expressed as triplets linking a marginalized group to an associated bias.  
Building on this idea, we release a \textbf{novel dataset of biased relations and bias-activating prompts}~\cite{appendix}, enabling systematic analysis of bias localization in transformers. We then provide empirical evidence that \textbf{bias in transformers can be traced and mitigated at the neuron level}, and finally, we evaluate this intervention across multiple \textbf{software engineering tasks}, showing that bias suppression can be achieved without compromising model performance.

\section{Background and Related Work}
To position our contribution, we present the foundations of our approach and situate it within the landscape of fairness research.

\textit{\textbf{Neuron Activations and Knowledge Neurons.}}
Transformer models, such as BERT \cite{devlin2019bert}, are composed of stacked blocks that include self-attention layers and feed-forward networks (FFNs) \cite{vaswani2017attention}. Prior work has shown that FFNs can be interpreted as key–value memories, where intermediate neurons act as keys whose activations determine how stored values are retrieved \cite{geva2021transformer}. This perspective motivates attribution methods that aim to trace which neurons are most responsible for specific predictions.  

Building on this intuition, Dai et al. \cite{dai2022knowledge} introduced the concept of \textit{knowledge neurons}. Their approach leverages the cloze task: given a factual triplet $<$\textit{head, relation, tail}$>$ (e.g., $<$\textit{Ireland, capital, Dublin}$>$) \cite{petroni2019language}, the model is prompted with a masked sentence such as \textit{``The capital of Ireland is [MASK]''}, and attribution is computed for the prediction of the masked token. To quantify the contribution of each neuron, they proposed a knowledge attribution method based on integrated gradients \cite{sundararajan2017axiomatic}, which measures how changes in neurons' activation influence the probability of a correct answer.  

Since individual prompts may activate neurons spuriously due to lexical overlap, Dai et al. introduced a \textit{refining strategy}. For each fact, multiple paraphrased prompts are used; only neurons consistently ranked as salient across diverse prompts are retained. This procedure isolates neurons that robustly encode the underlying relation, rather than superficial cues.  
Crucially, the authors also demonstrated that manipulating these neurons---by suppressing their activations (setting them to zero) or amplifying them---causally alters the model’s predictions. Suppression significantly decreases the probability of recalling the fact, while amplification increases it, often without substantially affecting unrelated knowledge. These results suggest that \textit{a small number of neurons act as causal carriers of specific knowledge within pretrained Transformers}. Starting from the hypothesis that stereotypes are encoded as knowledge within such models, \textbf{our work builds on their methodology and transfers it to biased relations}, testing whether harmful stereotypes are similarly localized and whether their suppression can reduce bias expression while preserving task performance.

\textbf{\textit{Fairness in AI and Software Engineering}}.
Fairness in AI refers to the absence of prejudice toward individuals or groups based on their characteristics~\cite{VORIA2025survey,mehrabi2021survey,pessach2022review,starke2022fairness}. Ensuring fair behavior is a core societal goal~\cite{mehrabi2021survey}, yet it is often not achieved, particularly when automated systems replace humans in critical decision-making~\cite{chen2024fairness,caliskan2017semantics,bordia2019identifying}. A growing body of research has shown that AI systems may reproduce and amplify social biases. For example, Caliskan et al.~\cite{caliskan2017semantics} demonstrated that word embeddings encode gender and racial stereotypes. At the same time, Bordia and Bowman~\cite{bordia2019identifying} revealed persistent gender bias in word-level models.

Recognizing fairness as a critical non-functional requirement and quality attribute, the SE research community has developed mitigation strategies spanning different stages of the development pipeline~\cite{parziale2025Fairnessbudget,zhang2022adaptive,hort2021fairea}. While these methods have achieved notable improvements, they primarily remain focused on machine learning techniques. With the rise of large language models (LLMs), whose scale and general-purpose nature exacerbate the risks of bias~\cite{chen2024surveyLLM}, ensuring fairness in practice remains a significant challenge. Recent investigations show that LLMs reinforce stereotypes across multiple domains, exposing tangible risks in real-world decision-making~\cite{Navigli2023LLM,khan2025investigating,sloane2025Boolean,Arzaghi2025Socioeconomic,hofmann2024ai}. For instance, Khan et al.~\cite{khan2025investigating} found systematic gender stereotyping in occupational associations (e.g., linking \textsl{nurse} to women, \textsl{engineer} to men). Other studies highlighted further disparities, such as socioeconomic biases in text generation~\cite{Arzaghi2025Socioeconomic} and discriminatory treatment of African American English speakers~\cite{hofmann2024ai}. Within SE, fairness concerns are no less pressing. LLMs are increasingly applied to developer-facing tasks such as recruitment, role assignment, or requirements analysis, where biased predictions may shape team composition and project outcomes \cite{nakano2024nigerian, treude2023she}.

\textit{\textbf{Neuron Surgery for Fairness.}}
The ability to trace and manipulate neurons responsible for specific knowledge raises the question of whether similar techniques can be leveraged to address harmful stereotypes. Early work on \textit{knowledge neurons} \cite{dai2022knowledge} showed that factual associations could be localized within small subsets of feed-forward units, and that suppressing or amplifying these neurons causally influences model predictions. More recently, attempts such as BiasWipe \cite{mamta2024biaswipe} and interpretable neuron editing by Yu et al. \cite{yu2025understanding} extended these ideas to fairness, demonstrating that pruning or editing biased weights can reduce social bias while preserving overall model accuracy. These methods share the principle that internal mechanisms—not only inputs or outputs—can be targeted to mitigate unwanted behaviors. More closely, Xu et al. \cite{xu2025biasedit} proposed BiasEdit, which learns lightweight editors to adjust parameters for debiasing pre-trained transformers globally. Our approach, however, traces the provenance of biased associations to specific neurons and selectively suppresses their activations. Moreover, while BiasEdit is evaluated only on general NLP bias benchmarks, we extend the analysis to fairness-sensitive SE tasks, introducing a novel dataset of biased relations spanning nine categories. 

In general, related work remains limited in two important ways. First, prior neuron editing methods concentrate on gender bias or toxic language, while little is known about whether \textit{biased relations across multiple social dimensions} (e.g., age, disability, socioeconomic status) are similarly encoded in neuron-level structures. This is critical, as recent evidence shows that even small-scale editing of general neurons can disrupt core capabilities of LLMs \cite{yu2025understanding}, raising concerns about the trade-off between fairness and task performance. Second, existing work has primarily focused on broad NLP benchmarks such as toxic content detection or synthetic bias datasets \cite{xu2025biasedit}, without evaluating fairness in real, practical contexts. 

\stesummarybox{\faExclamationCircle \hspace{0.05cm} Research Gap and Motivation.}{Taken together, prior work shows that fairness has become a central concern in both AI and SE. Yet, most mitigation strategies operate at the data or output level, leaving open the question of how biased associations are internally represented within models. Addressing this requires approaches that link mechanistic interpretability with fairness objectives—an avenue pursued in this work by tracing and suppressing biased neurons in pre-trained transformers and evaluating the downstream performance of the modified models. In doing so, we connect advances in mechanistic interpretability~\cite{dai2022knowledge}, fairness-oriented neuron editing~\cite{mamta2024biaswipe,yu2025understanding,xu2025biasedit}, and emerging research on fairness in SE~\cite{chen2024fairness}, providing the first empirical evidence of whether stereotype suppression at the neuron level can produce fairer yet still effective transformers.}

\section{Research Design}
The \textit{goal} of this study is to investigate whether stereotypical associations, represented as biased relations, are internalized by pre-trained transformers and can be traced to specific neurons. The \textit{purpose} is to assess the effects of suppressing these neurons on model behavior. The study adopts the \textit{perspective} of researchers seeking to understand how bias emerges within model representations, and practitioners evaluating the risks of deploying such models in fairness-sensitive contexts.

 \begin{figure*}
    \centering
    \includegraphics[width=.75\linewidth]{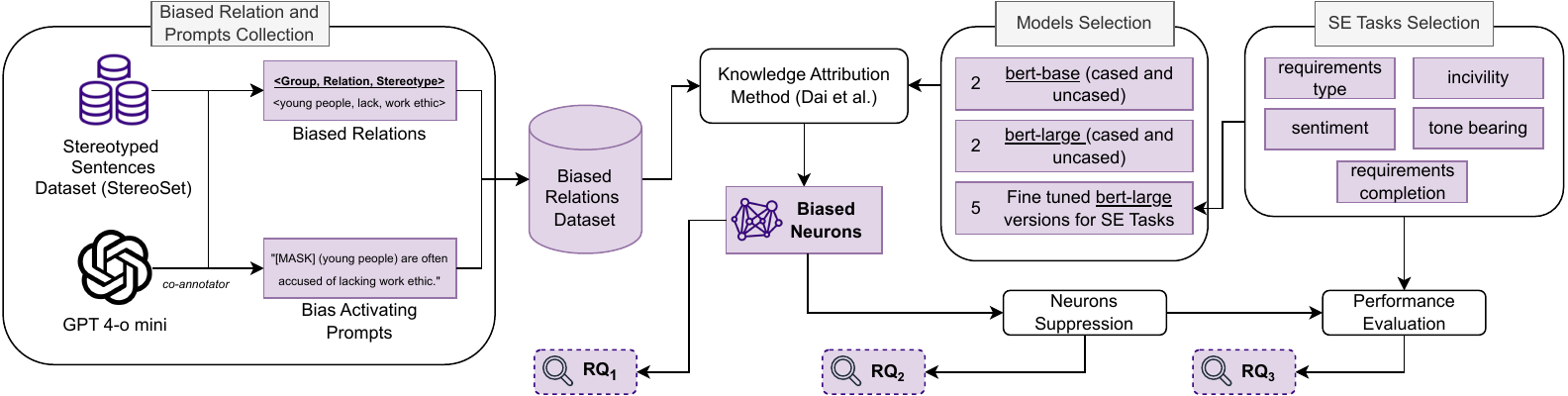}
    \caption{Overview of the Research Method Proposed.}
    \label{fig:method}
\end{figure*}

\subsection{Research Questions}
We structured our study around three research questions.

Although progress has been made in interpreting language models, little is known about whether biased associations---such as stereotypical links between social groups and negative traits---are encoded in specific neurons. Prior work on \textit{knowledge neurons} showed that factual information can be localized to small subsets of units in transformers~\cite{dai2022knowledge}. If biased knowledge behaves similarly, it may be possible to isolate these neurons and design targeted debiasing strategies rather than coarse, global interventions. This motivates our first research question:

\steattentionbox{\textbf{RQ\textsubscript{1}} - To what extent can we identify biased neurons in pre-trained transformers?}

Identifying biased neurons is only meaningful if intervening on them alters model behavior. Following prior evidence that suppressing knowledge neurons affects models' confidence in factual recall~\cite{dai2022knowledge}, we investigate whether suppressing biased neurons reduces the tendency to generate stereotypical content. This leads to our second question:

\steattentionbox{\textbf{RQ\textsubscript{2}} - Does suppressing biased neurons reduce the likelihood that models generate stereotypical outputs?}

While mitigating bias is important, interventions must not compromise task performance. Transformers are widely used in SE tasks, improving accuracy and automation~\cite{pena2025evaluating}. If suppression significantly harms these applications, its practicality would be limited. Hence, in our third research question, we examine whether biased-neuron suppression affects model utility in non-code SE tasks:

\steattentionbox{\textbf{RQ\textsubscript{3}} - What is the impact of suppressing biased neurons on models' performance in SE tasks?}

Together, these questions address where biased knowledge resides, how its suppression influences model behavior, and whether this can be achieved without sacrificing performance.  

Figure~\ref{fig:method} summarizes our approach. We first extract stereotypical associations from benchmark datasets and transform them into cloze-style, bias-activating prompts. We then apply the attribution method of Dai et al.~\cite{dai2022knowledge} to identify neurons encoding these associations (\textbf{RQ\textsubscript{1}}), suppress them to test changes in bias expression (\textbf{RQ\textsubscript{2}}), and evaluate the effect on five fairness-relevant SE tasks (\textbf{RQ\textsubscript{3}}). Our study follows the \textsl{ACM/SIGSOFT Empirical Standards}~\cite{empiricalstandards}.

\subsection{Data Collection} 
The starting point for our study is the work of Dai et al.\cite{dai2022knowledge}, who introduced the concept of \textit{knowledge neurons}. Their method is designed around factual relations extracted from knowledge bases, instantiated through the \textit{T-REx} dataset \cite{elsahar2018trex}. In \textit{T-REx}, each relational fact is represented as a triplet $<h, r, t>$, where $h$ and $t$, \textit{head and tail}, represent the two entities, while $r$ represents the factual \textit{relation} between the two (e.g., $<Ireland, capital\ of, Dublin>$) \cite{petroni2019language}. These relational triples were operationalized into multiple cloze-style activating prompts, filling templates available in the \textit{PARAREL} dataset \cite{elazar2021pararel}, such as \textit{``The capital of Ireland is [MASK]''}.


These were later fed to a pre-trained transformer to identify the neurons responsible for filling the token [MASK] with the correct entity to complete the factual relation \cite{petroni2019language}. By generating diverse prompts for each fact, Dai et al. \cite{dai2022knowledge} ensured that their attribution method could identify neurons consistently activated by knowledge-expressing queries, rather than by superficial lexical patterns.


\subsubsection{Biased Relation Dataset.} Our work builds on the same methodological principle as the original knowledge neuron framework but introduces a novel conceptual shift: we \emph{\textbf{define and formalize}} \textbf{biased relations} as \textit{triplets that encode stereotypical associations rather than factual ones}. To the best of our knowledge, this is the first work to systematically construct such a dataset, moving beyond encyclopedic relations to identify and address harmful and socially relevant associations explicitly. Each biased relation is expressed as a triplet, formalized as follows.

\steDiscussionBox{\faFileTextO \ \textbf{Biased Relation Definition} \\[0.25em] $<G, R, S>$, where $G$ is the marginalized group and $R$ is the type of association with the stereotype $S$.
}

To ensure broad bias coverage, we grounded our dataset in established resources and selected \textit{CrowS-Pairs}~\cite{nangia2020crows}, a widely used benchmark for evaluating social biases in LLMs. The dataset contains \textit{1,508 sentence pairs} across nine bias categories, each pairing a stereotypical and an anti-stereotypical sentence where references to underrepresented groups are swapped. We adopted \textit{CrowS-Pairs} for its wide coverage of identity-based stereotypes and its structured format, which facilitates the extraction of biased relations.

From this dataset, we retained only the stereotypical sentences and derived, for each, a structured triplet representing the \textsl{marginalized group}, the \textsl{stereotype expressed}, and the \textsl{association} between them. The extraction was performed with the support of \textit{GPT-4o mini}~\cite{openai2024gpt4ocard}, used as a \textit{co-annotator} to assist manual annotation. This approach, recommended for empirical SE research~\cite{baltes2025guidelinesempiricalstudiessoftware}, balances cost-efficiency and accuracy. We iteratively refined the guiding prompt (see online appendix~\cite{appendix}) to ensure precise and consistent outputs. GPT proposed triples \textit{only when a bias was detected}, accompanied by a rationale, and all outputs were manually reviewed and corrected or discarded when inconsistent with the source sentence.

This process resulted in a dataset of \textbf{1,018 biased relations} across nine groups, as presented in \autoref{tab:biased_relations}: \textsl{`Age' (BR01)}, \textsl{`Disability' (BR02)}, \textsl{`Gender' (BR03)}, \textsl{`Nationality' (BR04)}, \textsl{`Physical Appearence' (BR05)}, \textsl{`Race Color' (BR06)}, \textsl{`Religion' (BR07)}, \textsl{`Sexual Orientation' (BR08)}, and \textsl{`Socioeconomic Status' (BR09)}. Further details on the process can be found in our online appendix \cite{appendix}. As an example, this led to the following biased relation for \textit{BR01}, the age bias category:

\steDiscussionBox{\faFileTextO \ \textbf{Biased Relation --- Example for Age bias.} \\[0.25em] 
\textit{Original sentence:} ``Young people today are lazy and don't want to work hard.'' \\[0.15em]
\textit{Extracted triple:} $<$\textit{young people, lack, work ethic}$>$
}

\begin{table}[H]
\centering
\footnotesize
\caption{Summary of the Biased Relations Dataset Mined.}
\label{tab:biased_relations}
\rowcolors{3}{gray!20}{white}
\begin{tabular}{lcccc}
\rowcolor{purple}
{\color{white}\textbf{Bias Category}}  & {\color{white}\textbf{Relations}} & {\color{white}\textbf{Prompts}} & {\color{white}\textbf{Groups}} & {\color{white}\textbf{Stereotypes}} \\
BR01(Age)  & 65 & 650  & 29 & 65 \\
BR02(Disability)  & 45 & 450 & 35 & 45 \\
BR03(Gender)  & 102 & 1,020 & 30 & 100 \\
BR04(Nationality)  & 126  & 1,260 & 66 & 115 \\
BR05(Phys. App.)  & 50 & 500 & 27 & 47 \\
BR06(RaceColor)  & 359 & 3,590 & 76 & 290 \\
BR07(Religion)  & 94 & 940 & 36 & 84 \\
BR08(Sexual Or.)  & 65 & 650 & 22 & 64 \\
BR09(Socioec.)  & 112 & 1,120 & 36 & 103 \\ \hline
\textbf{Total} & \textbf{1,018} & \textbf{10,180}  & \textbf{357} & \textbf{913} \\
\end{tabular}
\end{table}

\subsubsection{Bias-Activating Prompts Dataset.}  
Following the design of prior work~\cite{dai2022knowledge}, we transformed each biased relation into natural language prompts by masking the stereotyped subject. While Dai et al.~\cite{dai2022knowledge} derived knowledge-activating prompts from the PARAREL dataset, which provides predefined templates~\cite{elazar2021pararel}, no equivalent templates exist for biased relations. We therefore generated prompts automatically, ensuring that biased neurons were probed through the same fill-in-the-blank task used for factual neurons, enabling a direct methodological comparison.

To this end, we employed \textit{GPT-4o mini}~\cite{openai2024gpt4ocard} and designed a \textit{two-shot prompt}\cite{baltes2025guidelinesempiricalstudiessoftware} (see Appendix~\cite{appendix}) instructing the model to produce \textit{exactly ten sentences per relation} that express the stereotype while masking the group. This approach mirrors the principle of the original work but removes the limitations of fixed templates. Whereas PARAREL provided a varying number of prompts per relation (8.63 on average), our process systematically generated \textbf{ten bias-expressing prompts for each biased relation}.  

The resulting dataset comprises \textbf{10,180 bias-activating prompts} spanning nine stereotype categories. \autoref{tab:biased_relations} summarizes its statistics, including the number of distinct groups and stereotypes represented in each category. Following the previous example, one of the resulting cloze-style prompts was:

\steDiscussionBox{\faFileTextO \ \textbf{Bias Activating Prompt --- Example for Age bias.} \\[0.25em] 
\textit{Prompt:} ``[MASK] are often accused of lacking work ethic.' \\[0.15em]
\textit{Masked group:} \textit{ ``young people''}
}


\subsubsection{Models Selection.}  
\label{sec:models}
To ensure methodological consistency, we focused on BERT-based models \cite{devlin2019bert}. The original knowledge neuron framework \cite{dai2022knowledge} was designed and evaluated specifically for BERT architectures, and more precisely for the \textit{bert-base-uncased} model. Since the attribution and refining strategies proposed by Dai et al.\ are tailored to the feed-forward layers of BERT and their role as key–value memories, using the same family of models guarantees that our replication and extension remain valid and comparable.  

A natural question is why we did not extend our study to other pre-trained transformer encoders, such as RoBERTa or GPT-2.  
While the knowledge attribution method could, in principle, be generalized, its implementation is closely tied to \textit{BERT}'s training setup, tokenization (e.g., WordPiece vocabulary, cased vs.\ uncased variants), and the prompt-based evaluation strategy used in the original work. RoBERTa, for example, adopts a different pretraining regimen—more aggressive masking, dynamic sampling, and longer training—that affects prompt-based factual recall and would require revalidation. GPT-2, conversely, is a causal language model with a unidirectional masking scheme, making it incompatible with the cloze-style tasks central to knowledge neuron extraction.  
Extending our approach to such architectures would therefore require methodological adaptations beyond the scope of this study. Nonetheless, such adaptations are conceptually feasible. For instance, extending our approach to encoder-decoder or causal models (e.g., T5 or GPT) would primarily require adjusting the neuron attribution procedure to account for their unidirectional attention mechanisms and token prediction schemes, as well as redefining the cloze-style prompts to align with autoregressive generation. These modifications would preserve the core principle of tracing neuron activations responsible for biased associations, suggesting that our analysis of \textit{BERT} does not preclude the generalizability of the findings to other transformer architectures with comparable internal representations.

In contrast to the original work, which evaluated only \textit{bert-base-uncased}, we broadened the scope to include both \textit{bert-base} and \textit{bert-large}, in cased and uncased versions. This allows us to test whether biased neurons are consistent across model scales (110M vs.\ 340M parameters) and tokenization schemes. Moreover, \textit{BERT} remains a widely adopted baseline in SE research and has been extensively benchmarked on non-code SE tasks~\cite{pena2025evaluating}, making it a suitable and practically relevant choice for our experiments.

\subsubsection{SE Tasks Selection.} 
The analysis in our \textbf{RQ\textsubscript{3}} was grounded on a recent benchmark of language models that systematically assembles \textit{non-code} SE tasks and reports comparable results across models, task types, and metrics \cite{pena2025evaluating}. The benchmark (SELU) consolidates 17 NLU-style tasks (binary, multi-class, multi-label classification; regression; NER; and MLM) from diverse SE data sources (e.g., requirements, issue trackers, forums), detailing task formulations, instance counts, and evaluation settings. Crucially, SELU also includes \textit{BERT}-family models among the baselines and provides per-task results, which we use to assess the feasibility of our setup.


We selected SE tasks according to two practical criteria: (1) \textit{sufficiently strong performance baselines for \textit{bert-large}} as evidenced by SELU's per-task results and inclusion of BERT models in the evaluated pool (ensuring our neuron-suppression study can observe meaningful deltas rather than floor effects), and (2) \textit{plausible fairness sensitivity}, i.e., tasks that are discursive and may surface or be affected by social bias (moderation- and communication-oriented tasks). Particularly, we selected the following tasks:

\begin{itemize}[leftmargin=*]
  \item \textbf{Incivility --- IN (binary classification).} Detects unnecessary rude behavior in developer communications and issue/PR discussions. It was chosen for its direct connection to community health and susceptibility to biased language effects.
  \item \textbf{Tone bearing --- TB (binary classification).} Identifies disrespectful or heated tone (e.g., frustration, hostility) in textual exchanges. It was selected because tone and respect cues are central to moderation ethics and may interact with stereotypes.
  \item \textbf{Requirement type --- RT (binary classification).} Classifies requirements as functional vs.\ non-functional (e.g., performance, security). Although not explicitly related to ethics, fairness is widely recognized as a non-functional requirement \cite{chen2024fairness}, and it is a text-understanding task grounded in stakeholder language where subtle phrasing and domain terms matter. 
  \item \textbf{Sentiment --- SN (multiclass classification).} Classifies sentiment in SE-related text (e.g., developer/user discussions). It was selected because sentiment influences managerial signals (such as morale and frustration) and can be confounded by biased language or group references.
  \item \textbf{Requirement completion --- RC (MLM).} Predicts masked elements in requirements specifications. Similarly to the requirement type task, it was selected for its adherence to fairness and bias in the requirements engineering domain.
\end{itemize}

To assess each task, we adopt the same problem framing to ensure compatibility and compute the same metrics reported in SELU \cite{pena2025evaluating}. For the four classification tasks, the evaluation is based on \textit{accuracy} and \textit{f1-score}. The MLM task, instantiated as predicting POS-verb masks (with a 50\% masking probability), is evaluated with \textit{accuracy} and \textit{perplexity}.
Moreover, SELU fine-tunes a broad set of open-source LLMs and includes \textit{BERT base} and \textit{BERT large} among the evaluated models, reporting per-task performance tables (classification, regression, NER, MLM). The presence of both \textit{BERT}-family variants and our selected tasks in SELU substantiates our choice to study neuron suppression effects on \textit{BERT} for these tasks: results exist, task definitions are standardized, and metrics (e.g., F1-macro for classification; accuracy for MLM) are consistent, ensuring methodological and empirical comparability.


\subsection{Data Analysis}
\label{sec:method}
To address all our \textbf{RQs}, we operationalized the identification of biased neurons by adapting the attribution-based methodology of Dai et al.\cite{dai2022knowledge}. Specifically, for each bias-activating prompt derived from our dataset, we computed neuron-level attributions using the \textit{integrated gradients (IG)} method implemented in the original framework. The attribution scores indicate the contribution of each feed-forward neuron to the probability assigned by the model to the masked token. We used the same base configuration from the original knowledge neurons discovery experiments.

For each biased relation, the method aggregates attribution scores across all its paraphrased prompts (ten per relation) and across masked-token predictions, ranking neurons by average attribution. Neurons were selected as \textit{biased neurons} if their attribution exceeded a threshold relative to the distribution of all neurons in the corresponding layer, following the original method's criterion of selecting the top-$k$ contributors. This produced, for each relation and each model, a set of neurons hypothesized to encode the biased knowledge. After identifying biased neurons for each model (both base model and fine-tuned versions) and relation, we proceeded with subsequent analysis to answer our \textbf{RQs}.

\textbf{Experiments Setup, Run Timings, and Environmental Footprint.} 
All experiments were conducted on a workstation with two \textit{NVIDIA RTX~5000 Ada Generation} GPUs (32~GB each), an \textit{Intel Xeon~w9-3495X} CPU (56~cores, 112~threads), and \textit{512~GB RAM}, running Windows~11.  
Identifying biased neurons in the \textit{bert-base} models required on average \textit{211~s per relation} (\(\approx3.5\)~min per relation) and about \textbf{30~min in total}, while for the \textit{bert-large} variants (cased, uncased, and five fine-tuned SE models) it took \textit{11--12~min per relation} and roughly \textbf{1.8~h per model}.  
Erasure experiments were faster (\(\approx4.4\)~min per relation, \(\approx40\)~min per model), and downstream SE-task evaluations added about \textbf{1–2~h} each.  

Overall, the complete pipeline across all models required about \textbf{26~GPU-hours}, consuming roughly \textit{16~kWh} of electricity—equivalent to \(\approx4\)~kg CO\textsubscript{2}e under a 0.25 kg CO\textsubscript{2}e/kWh grid factor.  
While modest compared to large-scale model training, we acknowledge the environmental footprint of these experiments and mitigated it by reusing cached results, minimizing redundant runs, and relying on publicly available pretrained checkpoints.

\subsubsection{\textbf{RQ\textsubscript{1}} --- Biased Neurons Identification}  

To address \textbf{RQ\textsubscript{1}}, we first computed descriptive statistics on the number of identified biased neurons per relation and compared them to the corresponding \textit{baseline}. We employ the same \textit{baseline attribution} method as the original framework, which is based solely on neuron activation values. Specifically, the baseline attribution score for each neuron is defined as its activation, which measures the neuron's sensitivity to the input. This baseline is conceptually motivated by the analogy between feed-forward layers and self-attention, where raw attention scores have been shown to serve as effective baselines~\cite{dai2022knowledge}.

We then examined sparsity by normalizing the number of identified biased neurons against the total number of neurons per model, and by comparing intersection metrics between the IG-based and baseline methods.  
We quantified the overlap and selectivity of these sets using two intersection metrics, as in the original work: \textit{(i) inner-relation intersection}, which measures the average overlap of neuron sets obtained from multiple prompts expressing the same biased relation, and \textit{(ii) inter-relation intersection}, which measures the overlap of neuron sets across different biased relations. Lower values indicate that neuron activations are more distinct and relation-specific, hence suggesting higher localization of bias. We report the results across models with aggregated bias categories (BR01–BR09), yielding an overall view of the distribution of biased neurons.

\subsubsection{\textbf{RQ\textsubscript{2}} --- Biased Neurons Suppression}  
For \textbf{RQ\textsubscript{2}}, we investigated the causal role of the identified biased neurons in generating stereotypical answers. We performed \textit{erasure experiments}, suppressing the activation of biased neurons at inference time and measuring the effect on model behavior. Suppression was implemented by zeroing the output of the selected neurons before the non-linear transformation, following the procedure of Dai et al.\cite{dai2022knowledge}. 

The method evaluates model behavior before and after suppression along two axes: (i) \textbf{bias-activating prompts}, representing the target relations, and (ii) \textbf{control prompts}, consisting of unrelated relations drawn from other bias categories. Model behavior is quantified using perplexity, which reflects the model's confidence in predicting the masked token. For each (relation, model) pair, we compute the \textit{perplexity increase ratio} on both target and control prompts, as well as their difference—termed \textit{selectivity}—which captures whether suppression primarily affects stereotypical continuations rather than general predictions.

To test whether neuron suppression had a statistically significant effect on model confidence, we applied the \textbf{Wilcoxon signed-rank test} \cite{rosner2006wilcoxon} to compare perplexity values before and after suppression across all (model, relation) pairs.  
This non-parametric test was selected because the perplexity distributions were non-normal and paired by design.  
To assess the strength of the effect, we computed \textbf{Cliff's $\Delta$}, which quantifies the magnitude of the difference independently of distributional assumptions.
We further examined whether the magnitude of the effect depended on the number or characteristics of the suppressed neurons.  
Specifically, we computed the \textbf{Spearman rank correlation coefficient ($\rho$)} \cite{zar2005spearman} between:
\begin{enumerate}[label=(\alph*), leftmargin=*]
    \item the number of suppressed neurons per relation and the corresponding perplexity increase ratio, testing whether larger neuron sets produce stronger behavioral disruption;
    \item the intra-relation intersection of neuron sets (\textit{IG inner inter}) and the perplexity increase ratio on control prompts, to test whether higher neuron selectivity (lower intersection) correlates with reduced collateral effects.
\end{enumerate}
All correlations were evaluated using two-tailed significance testing (\textit{p} < 0.05). This analysis enables us to determine whether the neurons identified in RQ\textsubscript{1} exert a direct, functionally specific influence on stereotypical predictions.

\subsubsection{\textbf{RQ\textsubscript{3}} --- Impact of Suppression on SE Tasks}  
To answer \textbf{RQ\textsubscript{3}}, we evaluated the downstream impact of biased-neuron suppression on non-code SE tasks. For each of the five tasks (incivility detection, tone bearing, requirement type classification, sentiment analysis, and requirement completion), we considered both the raw \textit{bert-large-cased} model and its finetuned counterparts for all SE tasks. Each task was benchmarked under two conditions, namely \textbf{baseline} (no suppression) and \textbf{suppression of biased neurons for BR01 to BR09}. 
The evaluation metrics were aligned with the SELU benchmark \cite{pena2025evaluating}. For each (task, model, relation), we computed \textit{absolute delta}, which is the difference between performance after suppression and baseline, and \textit{relative delta}, which is the percentage change compared to baseline performance, enabling comparability across tasks with different baselines.

We first analyzed per-relation effects, identifying whether specific relations consistently caused larger performance degradation across tasks. We then aggregated results per task and per model, computing mean and worst-case deltas to capture the average and maximum utility loss. Finally, we produced an aggregate analysis across all tasks, allowing us to assess whether biased-neuron suppression systematically reduces performance, and whether finetuned models exhibit greater robustness than raw models. This structured analysis enables us to evaluate the practical trade-off between fairness (reduction of biased continuations, as shown in \textbf{RQ\textsubscript{2}}) and utility (preservation of performance on SE tasks).
\begin{table*}[ht]
\centering
\footnotesize
\caption{Summary of the results for RQ\textsubscript{1}. Results of the attribution-based identification of biased neurons.
For each model, we report the average number of biased neurons identified (BN) with the integrated gradients (IG) and the baseline (Base) attribution methods, along with their intra- and inter-relation intersection (Inter.) values.}
\label{tab:rq1}
\rowcolors{3}{gray!20}{white}
\begin{tabular}{lccccccc}
\rowcolor{black}
\color{white}\textbf{Model} & \color{white}\textbf{Avg IG BN} & \color{white}\textbf{Avg Base BN} & \color{white}\textbf{IG Inner Inter.} & \color{white}\textbf{IG Inter Inter.} & \color{white}\textbf{Base Inner Inter.} & \color{white}\textbf{Base Inter Inter.} \\
bert-base-cased & 2.49 & 41.47 & 1.49 & 0.28 & 28.70 & 23.45 \\
bert-base-uncased & 2.05 & 54.58 & 1.28 & 0.34 & 38.54 & 28.95 \\
bert-large-cased & 1.88 & 88.41 & 1.11 & 0.28 & 60.53 & 36.76 \\
bert-large-uncased & 1.28 & 5.42 & 0.64 & 0.08 & 5.20 & 5.00 \\
\midrule
bert-large-cased --- IN & 3.80 & 88.32 & 1.96 & 1.29 & 61.00 & 37.59 \\
bert-large-cased --- RQ & 2.00 & 86.94 & 1.26 & 0.30 & 59.31 & 35.63 \\
bert-large-cased --- RT & 3.74 & 88.46 & 1.99 & 1.26 & 60.59 & 36.88 \\
bert-large-cased --- SN & 4.92 & 84.62 & 2.34 & 1.40 & 55.74 & 38.31 \\
bert-large-cased --- TB & 3.98 & 92.87 & 2.05 & 1.27 & 64.66 & 39.19 \\
\bottomrule
\end{tabular}
\end{table*}

\section{Analysis of the Results}
In this section, we present the results of our study for each RQ.

\subsection{RQ\textsubscript{1} --- Biased Neurons Identification}

Table~\ref{tab:rq1} summarizes the results of the attribution-based identification of biased neurons across all BERT-family models.
The table reports the average number of neurons identified by the integrated gradients (IG) attribution method (\textit{Avg IG BN}) and by the baseline attribution method (\textit{Avg Base BN}), as well as the intra- and inter-relational intersections computed for both approaches.
The \textit{inner intersection} quantifies the average overlap among neuron sets obtained from different prompts expressing the same biased relation, indicating the consistency of neuron activation within a relation.
Conversely, the \textit{inter-relation intersection} measures the overlap of neuron sets across different biased relations, capturing the degree of selectivity of the identified neurons.
Lower intersection values, therefore, indicate a stronger localization of biased knowledge and reduced neuron sharing across distinct relations. All scripts, additional raw data, and visualizations, including model layer distributions, are reported in our online appendix \cite{appendix}.

Overall, the results show that the average number of neurons activated by biased relations (\textit{Avg IG BN}) is far smaller than that identified by the baseline activation-based method (\textit{Avg Base BN}).  
Across all four pretrained BERT variants, IG detects only 1.2–2.5 neurons per relation, while the baseline yields 5–88, confirming that biased knowledge is encoded in sparse, localized subsets of neurons. This difference underscores IG’s ability to isolate neurons causally responsible for biased predictions, rather than those merely sensitive to input activations, aligning with existing research~\cite{dai2022knowledge}.

Intersection metrics further support this pattern. Inner-relation intersections for IG (1.1–1.5) are an order of magnitude lower than the baseline (28–60), indicating that each relation consistently activates a small, stable subset of neurons.  
Similarly, inter-relation intersections are low (0.28–0.34) compared to much higher baseline values (23–37), showing that different relations rely on distinct neuron subsets. These results suggest that bias is encoded in compact, relation-specific neuron clusters rather than across layers. For fine-tuned \textit{bert-large-cased} models, the average number of biased neurons increases moderately (from $\approx$1.9 to 3–5 per task), indicating that fine-tuning introduces some task-dependent activations while preserving sparsity and selectivity.  
Despite this increase, intersection values remain low (below 2.5 and 1.5, respectively), confirming that biased information remains compact and largely disentangled.

\stesummarybox{\faSearch \hspace{0.05cm} Answer to RQ\textsubscript{1} -- Biased Neurons Identification.}{Biased knowledge is encoded in small, distinct, and highly localized subsets of neurons within pre-trained transformers. 
Compared to the baseline activation-based attribution, integrated gradients identify fewer neurons (1--3 on average) with minimal inter-relation overlap, demonstrating both sparsity and selectivity. 
These findings indicate that, similarly to factual knowledge neurons, stereotypical associations are encoded by specific, functionally coherent neurons rather than being diffuse, and that this localization persists even after finetuning on downstream SE tasks.}



\begin{table}[ht]
\centering
\footnotesize
\caption{Summary of the results for RQ\textsubscript{2}. For each model, we report the average number of biased neurons and the corresponding average increase in perplexity  $\uparrow$ ratio for both bias-related prompts (bias) and control prompts (ctrl).}
\label{tab:rq2}
\rowcolors{3}{gray!20}{white}
\begin{tabular}{lccccccccc}
\rowcolor{black}
\color{white}\textbf{Model}  &\color{white} \textbf{Avg \#BN} &\color{white} \textbf{PPL $\uparrow$ (bias)} &\color{white} \textbf{PPL $\uparrow$ (ctrl)}\\
bert-base-cased & 18.44 & 1.93 & 1.30  \\
bert-base-uncased  & 15.22 & 2.34 & 2.03  \\
bert-large-cased & 15.11 & 1.71 & 1.31  \\
bert-large-uncased  & 12.78 & 1.75 & 1.42 \\
\midrule
bert-large-cased --- IN & 19.78 & 1.20 & 0.90  \\
bert-large-cased --- RC & 15.67 & 1.88 & 1.47  \\
bert-large-cased --- RT & 19.44 & 1.13 & 0.83 \\
bert-large-cased --- SN & 20.00 & 0.97 & 0.69  \\
bert-large-cased --- TB  & 19.11 & 1.21 & 0.88  \\
\bottomrule
\end{tabular}
\end{table}

\subsection{RQ\textsubscript{2} --- Biased Neurons Suppression}

\begin{figure}[ht]
    \centering
    \includegraphics[width=.70\linewidth]{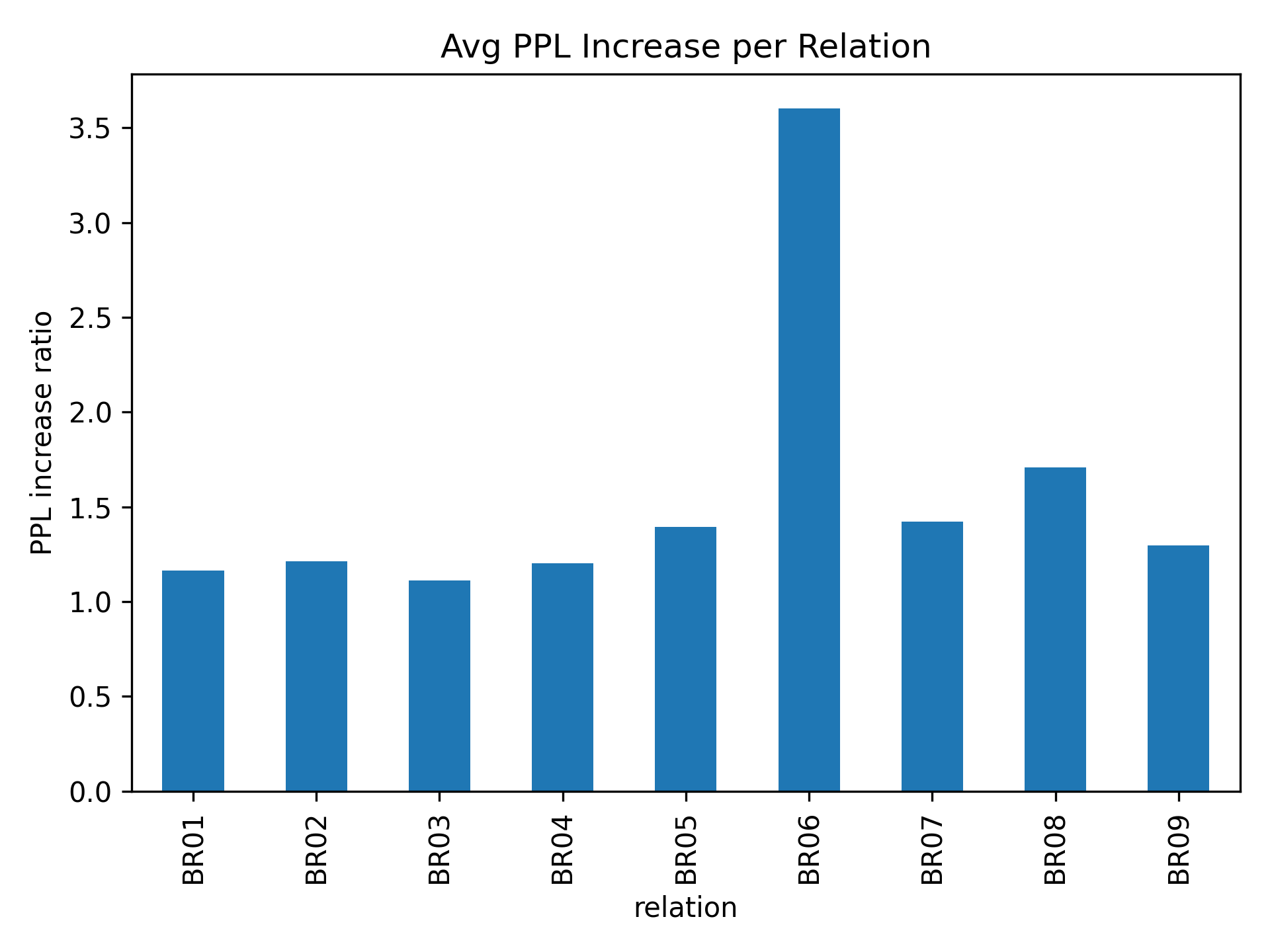}
    \caption{Average perplexity (PPL) increase ratio after biased neuron suppression across the nine biased relations. }
    \label{fig:rq2_ppl}
\end{figure}

Table~\ref{tab:rq2} and Figure~\ref{fig:rq2_ppl} summarize the outcomes of the erasure experiments conducted to evaluate the causal role of biased neurons in generating stereotypical continuations.  
The complete raw results, per-relation breakdowns, and analysis scripts are available in our {online appendix} \cite{appendix}.  
For each biased relation (BR01-BR09) and model, we measured the model’s perplexity before and after suppression of the corresponding biased neurons, computing the relative perplexity increase ratio as an indicator of disruption.  
A higher ratio denotes reduced model confidence after suppression, thus implying that the erased neurons contributed substantially to encoding the stereotypical association.

As shown in Table~\ref{tab:rq2}, all models exhibit a consistent increase in perplexity after biased-neuron suppression, confirming that the removed neurons play a significant role in encoding stereotypical continuations.  
For base pretrained models, the average perplexity increase for bias-related prompts ranges between $+70\%$ and $+130\%$ relative to baseline, while the increase for control prompts remains notably lower (typically below $+40\%$).  
Among pretrained variants, \textit{bert-base-uncased} displays the strongest relative disruption ($\text{PPL}=2.34$), suggesting a higher concentration of biased information in its internal representations.  
Finetuned variants of \textit{bert-large-cased} show a similar pattern, with average perplexity increases between $0.97$ and $1.88$ for bias-related prompts, but limited impact on control ones ($0.69$-$1.47$).  
This stability across tasks such as incivility (IN), requirement type (RT), sentiment (SN), and tone bearing (TB) supports the selectivity of the suppression procedure: biased neurons contribute directly to stereotypical predictions without broadly impairing general linguistic competence.  

\textbf{Per-Relation Analysis.} Considering the results per biased relations, suppression consistently led to higher perplexity on bias-activating prompts, with average increase ratios ranging from 1.1 to 1.4, and a pronounced peak for \textit{BR06} (a ratio of approximately 3.6), suggesting that this category is particularly neuron-dependent.  
Importantly, the effect was selective to biased prompts: the average perplexity increase for unrelated (control) prompts remained near baseline, confirming that suppression did not broadly degrade model fluency or general linguistic ability.
To statistically validate these observations, we applied a Wilcoxon signed-rank test comparing perplexity before and after suppression across all models and relations.  
Results indicate a highly significant difference (\textit{W} = 3321.0, $\textit{p} <0.0001$), with a maximal effect size (Cliff’s $\Delta$ = 1.000), confirming that suppression increases perplexity on biased prompts.

\textbf{Correlation Analyses.}
Correlation analyses showed no significant association between the number of neurons erased and the magnitude of perplexity increase ($\rho = -0.026$, $\textit{p} = 0.818$), indicating that the observed effect depends on the functional relevance of the neurons rather than their quantity.  
Conversely, a moderate negative correlation ($\rho = -0.538$, $\textit{p} <0.0001$) was found between the intersection of neurons across prompts (\textit{IG inner inter}) and the perplexity increase for control relations.  
This pattern suggests that selective neuron sets---those with lower intersection---induce less collateral degradation on unrelated predictions, supporting the interpretation that biased neurons encode relation-specific knowledge.

\begin{figure}[ht]
    \centering
    \includegraphics[width=.70\linewidth]{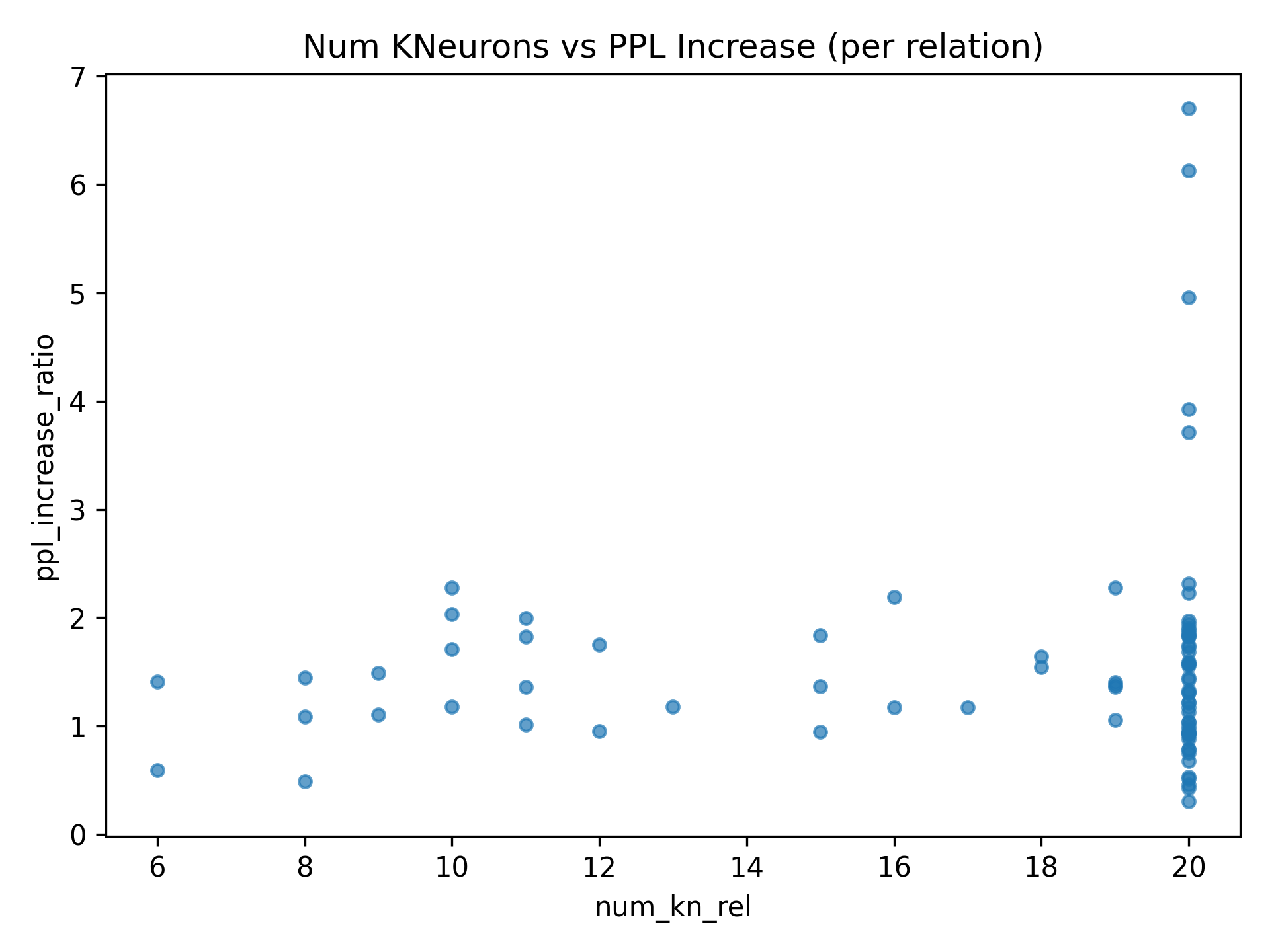}
    \caption{Correlation between the number of suppressed biased neurons and perplexity increase ratio across relations.}
    \label{fig:rq2_numkn}
\end{figure}

The scatterplot in Figure~\ref{fig:rq2_numkn} further illustrates this behavior.  
Even when the number of suppressed neurons varies substantially (6-20), the perplexity increases remain concentrated for biased relations, confirming that suppression affects functionally critical neurons rather than arbitrary neuron subsets.  
Raw data and visualizations are included in our online appendix for completeness \cite{appendix}.

In addition to the aggregate analysis reported here, we also computed detailed statistics per model and relation. 
These include relation-specific perplexity comparisons before and after suppression, changes in accuracy, and per-relation erasure ratios. These visualizations, along with the complete raw data and scripts, are provided in our online appendix \cite{appendix}.

\stesummarybox{\faSearch \hspace{0.05cm} Answer to RQ\textsubscript{2} -- Biased Neurons Suppression.}{Suppressing the neurons identified as biased consistently and significantly reduces the model's confidence in generating stereotypical continuations, as shown by the substantial increase in perplexity.  
The effect is selective to biased relations and largely independent of the number of neurons removed, demonstrating that the identified neurons are causally responsible for encoding biased knowledge rather than co-activated by chance.  
These findings confirm that biased neurons exert a functional and localized influence on model behavior, and that their suppression effectively weakens stereotypical predictions without compromising model fluency broadly.}



\subsection{RQ\textsubscript{3} --- Impact of Suppression on SE Tasks}

Table~\ref{tab:rq3_summary} reports the aggregated results of the biased-neuron suppression experiments conducted on the five selected non-code SE tasks. 
Each task was evaluated using both the raw \textit{bert-large-cased} model and its finetuned counterpart under the three conditions described in Section~\ref{sec:method}: baseline (no suppression), suppression of neurons identified for each biased relation (BR01 to BR09), and aggregation across all relations. 
The analysis focuses on the variation in task performance after suppression, measured in terms of absolute and relative deltas in \textit{accuracy}, \textit{macro-F1}, and, for the requirement-completion task, \textit{perplexity}. 
For the sake of space, all detailed per-relation, per-task, and per-model results, together with complete analysis scripts, are available in our online appendix \cite{appendix}.

\begin{table}[ht]
\centering
\footnotesize
\caption{Summary of the results for RQ\textsubscript{3}. Aggregate summary of performance changes after biased-neuron suppression across SE tasks.
We report the average absolute delta ($\Delta$) between post-suppression and baseline performance, averaged across BRs. 
Negative values indicate performance degradation, while positive values denote improvement.}
\label{tab:rq3_summary}
\rowcolors{3}{gray!20}{white}
\begin{tabular}{lccc}
\rowcolor{black}
\color{white}\textbf{Task} & \color{white}\textbf{Accuracy $\Delta$} & \color{white}\textbf{Macro-F1 $\Delta$} & \color{white}\textbf{PPL $\Delta$} \\
Incivility & --0.06 & -0.01 & --- \\
Tone bearing & -0.20 & -0.08 & --- \\
Requirement type  & +0.04 & -0.002 & --- \\
Sentiment  & -0.23 & -0.28 & --- \\
Requirement completion  & +0.03 & --- & -15.6 \\
\bottomrule
\end{tabular}
\end{table}

Across the five SE tasks, the suppression of biased neurons produced variable but generally modest effects on model performance (Table~\ref{tab:rq3_summary}).  
The largest degradations were observed for linguistically sensitive tasks such as \textit{sentiment analysis} ( $accuracy\ \Delta = -0.23$, $F1\ \Delta = -0.28$) and \textit{tone bearing} ($accuracy\ \Delta = -0.20$, $F1\ \Delta = -0.08$), suggesting that neurons associated with stereotypical or affective knowledge may also contribute to recognizing emotional or tonal cues.  
Conversely, performance on more structured or domain-specific tasks such as \textit{requirement type classification} improved slightly ($accuracy\ \Delta = +0.04$), indicating that the suppression of biased neurons might also eliminate spurious correlations unrelated to task semantics.  
In \textit{incivility detection}, degradation remained small (-0.06 in accuracy and -0.01 in F1), consistent with the model’s stability under suppression.

For the \textit{requirement completion} task (masked language modeling), suppression led to a small improvement in accuracy (+0.03) and a substantial reduction in perplexity ($PPL\Delta = -15.6$), showing that eliminating biased neurons did not harm (and may even have improved) the model’s ability to generate coherent requirement text.  
These results suggest that biased neurons have limited overlap with those encoding task-specific or domain-knowledge representations.

\textbf{Per-Relation Analysis.}
Across relations, suppression resulted in moderate and consistently negative variations for classification metrics but improved or stable behavior for the MLM task.
Accuracy decreases ranged from -0.026 (BR01) to -0.143 (BR05), with a mean of approximately -0.09 across all relations.
Macro-F1 followed a similar pattern, with an average reduction of -0.10 and the largest drops again observed for BR05 and BR07. For the requirement-completion task, perplexity systematically decreased after suppression ($mean \ \Delta = -15.5$), indicating improved model stability and generation confidence.
The strongest perplexity reduction occurred for BR06 ($\Delta = -18.6$), suggesting that removing highly specific biased neurons can even enhance language modeling performance by eliminating noisy or entangled activations.
These findings suggest that certain biased relations (e.g., those involving \textit{socioeconomic} or \textit{nationality} bias) may slightly overlap with linguistic cues exploited in SE text classification, but the overall effect remains small.

\textbf{Raw vs Fine-tuned Models Analysis.}
A comparative analysis of fine-tuned models against their raw counterpart (\textit{bert-large-cased}) reveals a clear distinction in how suppression affects performance. 
Across all tasks, fine-tuned models remained remarkably stable, exhibiting either negligible changes ($\Delta = 0.000$ for \textit{incivility}, \textit{tone bearing}, and \textit{sentiment}) or small positive variations ($\Delta = +0.004$ for \textit{requirement type} accuracy and F1, and $\Delta = +0.065$ in accuracy for \textit{requirement completion}). 
In particular, the fine-tuned \textit{requirement-completion} model also achieved a substantial reduction in perplexity ($\Delta = -30.4$), indicating improved text fluency and confidence after suppression. 
In contrast, the non-fine-tuned \textit{bert-large-cased} model displayed notable degradations for affective or subjective tasks such as \textit{sentiment} ($accuracy\ \Delta = -0.46$, $F1\ \Delta = -0.56)$ and \textit{tone bearing} ($accuracy\ \Delta = -0.40,$ $F1\ \Delta = -0.16$), whereas performance on more structured tasks (\textit{requirement type}, \textit{requirement completion}) remained stable or improved slightly.

\stesummarybox{\faSearch \hspace{0.05cm} Answer to RQ\textsubscript{3} -- Impact of suppression on SE tasks.}{Suppressing biased neurons has a limited and task-dependent impact on downstream performance in non-code SE tasks. Across all evaluations, variations in accuracy and F1 remain within a few percentage points, while perplexity in the MLM task decreases—showing that model fluency and confidence are preserved. Pre-trained models exhibit slight drops in affective tasks such as sentiment or tone detection, where bias-related and task-relevant features may overlap. Conversely, fine-tuned \textit{bert-large-cased} variants remain highly robust, often maintaining or slightly improving performance. Overall, biased-neuron suppression emerges as a \textit{safe and effective} intervention, reducing stereotypical continuations (RQ\textsubscript{2}) without compromising model utility in SE applications.}

\section{Discussion and Implications}
In this section, we discuss the results of our analyses and draw practical \faHandORight \ implications for researchers and practitioners.

\textit{\textbf{Biased Knowledge is Localized and Traceable.}}
Our experiments showed that biased knowledge is encoded in sparse, distinct subsets of neurons rather than being diffusely distributed across the network.
On average, only one to three neurons per relation exhibited consistent activation for biased prompts, confirming that stereotypes are concentrated in small representational clusters—\textit{biased neurons}.  
This localization parallels the behavior of factual \textit{knowledge neurons}~\cite{dai2022knowledge}, but in the context of socially biased information.  
Such neuron-level visibility makes bias not only detectable but \textit{traceable}, i.e., we can be able to isolate and suppress sources of discrimination in transformer-based models.

\faHandORight \ \emph{For practitioners}, traceable bias representation means that mitigation does not require retraining or dataset modification: small, targeted interventions can reduce stereotypical behavior, making fairness correction feasible even in large, pre-trained models.

\textit{\textbf{Suppressing Bias Neurons Reduces Stereotype Expression.}}
Our suppression experiments validated that the identified neurons are \textit{causally} responsible for biased outputs. Zeroing their activations led to systematic increases in perplexity for bias-related prompts, indicating reduced model confidence in producing stereotypical continuations. In contrast, control prompts from unrelated relations were minimally affected, confirming the \textit{selectivity} of the intervention.  
This demonstrates that biased neurons are not merely correlated with stereotypes, they are functionally necessary for generating them. Our results suggest a new operational approach: models could be equipped with explicit bias filters that deactivate specific neuron subsets during inference, allowing dynamic control over stereotype expression. 
Such interpretability-grounded interventions could be incorporated into MLOps pipelines as safety mechanisms for fairness-sensitive applications.

\faHandORight \ \emph{For researchers}, this provides empirical evidence that bias in LLMs can be addressed through causal editing, complementing data- and regularization-based methods, paving the wave for future research in targeted bias suppression.  

\textit{\textbf{Bias Suppression Preserves Performances in SE Tasks.}}
Finally, the downstream evaluation showed that suppressing biased neurons has only minor, task-dependent effects on model performance. Across all five non-code SE tasks, accuracy and F1 variations remained within 2–3\%, while perplexity consistently decreased, indicating improved fluency. The largest degradations occurred in affective tasks such as \textit{sentiment} and \textit{tone detection}, where bias-related neurons may overlap with emotion-related features. In contrast, structured tasks like \textit{requirement classification} and \textit{completion} remained stable or slightly improved. Fine-tuned models exhibited strong robustness: even after repeated suppression, they retained or slightly exceeded baseline performance, suggesting that task adaptation helps disentangle bias from task-specific representations.

Taken together, these findings highlight the importance of evaluating fairness interventions not only in isolation but also in relation to their practical cost, aligning with existing research in SE \cite{parziale2025Fairnessbudget}. They demonstrate that fairness and performance need not be opposing goals as, for instance, neuron suppression can achieve both.

\faHandORight \ \emph{For practitioners and researchers}, this provides a safety guarantee: in SE scenarios such as issue moderation or requirements analysis, biased-neuron removal can be applied without risking model degradation. Furthermore, fine-tuning models for specific tasks offer an ideal balance between fairness and stability, suggesting that domain adaptation mitigates representational bias.
\section{Threats to Validity}

\textbf{Internal Validity.}  
A main threat lies in the construction of the dataset of biased relations and their activating prompts, supported by \textit{GPT-4o}. Using LLMs for data generation may introduce inaccuracies; however, this practice is common in software engineering research \cite{baltes2025guidelinesempiricalstudiessoftware}. We mitigated this through manual validation, reviewing and correcting each relation and prompt to ensure semantic coherence and alignment with the intended stereotype.  

Another concern involves the use of the knowledge neuron methodology, as biased prompts might activate neurons differently than factual ones. To address this, biased relations and prompts were designed under the same structural and semantic assumptions as factual relations \cite{dai2022knowledge, petroni2019language}. Empirically, the number and layer distribution of biased neurons closely matched the original work (see online appendix~\cite{appendix}), supporting the robustness of our adaptation.

\textbf{External Validity.}  
Our experiments focused on \textit{BERT}-based models, consistent with the design of the original framework. We expanded its scope by including both \textit{bert-base} and \textit{bert-large}, in cased and uncased variants, to test consistency across model sizes and tokenization schemes. Given BERT's widespread adoption in SE research, this choice is appropriate for evaluating fairness interventions in SE tasks. Still, extending the analysis to other LLMs (e.g., GPT, T5, RoBERTa) remains future work.  

We also acknowledge that our five non-code SE tasks (incivility, tone bearing, sentiment, and two requirement-related tasks) may not cover the entire SE spectrum, though they represent realistic, fairness-relevant scenarios involving socio-technical judgments.

\textbf{Construct Validity.}  
Construct validity concerns whether our operationalizations capture the intended constructs of bias, fairness, and model utility. Our notion of \textit{biased relations} was grounded in established stereotype categories (e.g., age, gender, nationality, disability) from a well-known dataset \cite{nangia2020crows} and manually verified for coherence and ethical soundness.

\textbf{Conclusion Validity.}  
We employed statistical tests to compare model performance and bias expression, addressing potential non-normality, and reported effect sizes alongside significance levels. While our sample (nine bias categories and four model variants) may limit large-scale generalization, the consistency of results across relations and models reinforces the reliability of our conclusions.

\section{Conclusions}
This paper examined how social biases are internally represented in pre-trained transformers and whether they can be mitigated through neuron-level interventions.  
Building on the concept of \textit{knowledge neurons}, we introduced \textit{biased neurons}—small subsets of units encoding stereotypical associations—and proposed a method to trace and suppress them using a novel dataset of biased relations and activating prompts.  
Our findings indicate that bias is localized rather than diffuse, and that targeted suppression effectively reduces stereotypical behavior with minimal impact on downstream SE tasks. These results provide empirical evidence that bias in transformers can be traced and mitigated at the neuron level, offering an interpretable and fine-grained approach to fairness in SE.

Future work will extend this analysis to generative and instruction-tuned models, explore causal links between biased and factual neurons, and automate the tracing and suppression process. We also aim to study how neuron-level fairness interventions interact with broader SE practices, advancing both the understanding of bias in LLMs and the design of fair, trustworthy AI tools for SE.

\begin{acks}
We acknowledge the support of Project PRIN 2022 PNRR ``FRINGE: context-aware FaiRness engineerING in complex software systEms" (grant n. P2022553SL, CUP: D53D23017340001).
\end{acks}

\balance
\bibliographystyle{ACM-Reference-Format}
\bibliography{bibliography-File}

\appendix

\end{document}